\magnification=\magstep1
\tolerance=500
\bigskip
\rightline{24 July, 2016}
\bigskip
\centerline{\bf An Underlying Geometrical Manifold for Hamiltonian Mechanics}

\bigskip

\centerline{\it L.P. Horwitz${}^{1,2,3}$, A. Yahalom${}^4$, J. Levitan${}^3$
and M. Lewkowicz${}^3$}

\bigskip

\centerline{${}^1$ School of Physics, Tel Aviv University, Ramat Aviv
69978, Israel}
\centerline{${}^2$ Department of Physics, Bar Ilan University, Ramat Gan
52900, Israel}
\centerline{${}^3$ Department of Physics, Ariel University, Ariel 40700, Israel}
\centerline{${}^4$ Department of Electrical and Electronic Engineering,
Ariel University, Ariel 40700, Israel}

\bigskip
\noindent{\it Abstract}: We show that there exists an underlying
manifold with a conformal metric and compatible connection form, and a
metric type Hamiltonian (which we call the {\it geometrical picture}), that can be put into correspondence with the usual Hamilton-Lagrange mechanics. The requirement of dynamical equivalence of the two types of Hamiltonians, that the momenta generated by the two pictures be equal for all times, is sufficient to determine an expansion of the conformal factor, defined on the geometrical coordinate representation, in its domain of analyticity
with coefficients to all orders determined by functions of
the potential of the Hamiltonian-Lagrange picture, defined on the Hamilton-Lagrange coordinate representation,  and its derivatives. Conversely, if the conformal function is known, the potential of a
Hamilton-Lagrange picture can be determined in a similar way.  We show that arbitrary local variations of the orbits in the Hamilton-Lagrange picture can be generated by variations along geodesics in the geometrical picture and establish a correspondence which provides a basis for understanding how the instability in the geometrical picture is manifested in the instability of the the original Hamiltonian motion.
\smallskip
{\obeylines \smallskip
emails:
      ${}^1$larry@post.tau.ac.il
      ${}^2$ asya@ariel.ac.il
      ${}^3$ levitan@ariel.ac.il
      ${}^4$ lewkow@ariel.ac.il \smallskip}

\bigskip
\noindent{\bf 1. Introduction}
\smallskip
\par It has recently been shown [1] that the orbits generated by a
Hamiltonian of the form (the summation convention is used throughout)
$$ H= \delta_{ij} {p^i p^j \over 2m} + V(q), \eqno(1.1)$$
generating motions in the variables $\{ q,p\}$,
can be put into correspondence with the motions generated by a
Hamiltonian of the form (with $g_{ij}(x)$ invertible at every point, {\it i.e.}, no null space in momenta)
$$ H_G = g_{ij}(x) {\pi^i \pi^j \over 2m}, \eqno(1.2)$$
where we denote by $\{x\}$ the coordinates of the geometrical manifold (which we shall
call Gutzwiller coordinates; see, e.g., [2],[3])
on which $(1.2)$ generates motions
through the Hamilton equations for the canonical variables
$\{x,\pi\}$.
\par Eq. $(1.2)$ implies a geodesic motion associated with the metric $g_{ij}$.  The stability analysis of this representation of the dynamics, by means of geodesic deviation, has been shown to be a very sensitive indication of the local stability of the original Hamiltonian $(1.1)$ [1], and in many cases well-correlated with the chaotic behavior of the system [4]. In application to the restricted three-body problem [5][6], the method was shown to be highly effective.  In this case, small regions of instability did not affect the (stable) results, since their effects were suppressed by the ``uncertainty relation'' [7].  It has, furthermore, been shown that the instability associated with the geodesic deviation can be quantized and may have the interpretation of giving rise to excitations of the medium [8].
\par  A relation between the Hamiltonians $(1.1)$ and $(1.2)$
may be established by assuming [1] a conformal metric
$$ g_{ij}(x)= \phi(x)\delta_{ij}. \eqno(1.3)$$
We shall impose the condition of dynamical equivalence of the two Hamiltonians by requiring that  the momenta $p^i$ and $\pi^i$ be identical in the two systems for all $t$. It then follows that there  is a correspondence between the variables
$x$ and $q$ and the functions $\phi(x)$ and $V(q)$ such that
(both $H$ and $H_G$ are conserved, and can be assigned values $E$ and $E'$; the results are not affected by this freedom, since it only involves a constant shift in the potential $V$, and we take for convenience $H=H_G=E$)
$$\phi(x) = {E \over E-V(q)} \equiv F(q), \eqno(1.4)$$
defining the function $F(q)$, which we shall use for brevity in the following,
on the energy surfaces $H=H_G = E$.
\par In this paper, we show that the functions $\phi(x)$ and $F(q)$ on each of the manifolds can be put into correspondence through series expansion, and furthermore that the Poisson brackets and Lagrangians in both pictures are essentially equivalent, with variations in the geometrical picture driving variations in the Lagrangian of the Lagrange-Hamilton dynamics. We  may therefore  think of the geometrical picture in terms of an underlying geometry for the standard Langrange-Hamilton mechanics.
\par The Hamilton equations
$$ {\dot x}_i = {\partial H_G \over \partial \pi^i}\ \ \ \ {\dot \pi}_i =- {\partial H_G \over \partial x_i} , \eqno(1.5)$$
 applied to the Gutzwiller Hamiltonian $(1.2)$,
provide the relations
$${\dot x}_i = {1 \over m} g_{ij} \pi^j \eqno(1.6)$$
and
$$ \eqalign{{\dot \pi}^j &=- {1 \over 2m} {\partial g_{k\ell} \over \partial x_j } \pi^k \pi^\ell \cr
&= - {1 \over 2m}  {\partial g_{k\ell} \over \partial x_j } m^2 g^{km} g^{\ell n} {\dot x}_m {\dot x}_n \cr
&= { m \over 2}  {\partial g^{mn} \over \partial x_j }{\dot x}_m {\dot x}_n \cr}, \eqno(1.7)$$
where we have used the identity
$$ {\partial g^{ij} \over \partial x_m}= - g^{ik} {\partial g_{k\ell} \over
\partial x_m}g^{\ell j}. \eqno(1.8)$$
Then since, from $(1.6)$, we have
$$ {\ddot x}_i = {1\over m } {\partial g_{ij} \over \partial x_m} {\dot x}_m\pi^j + { 1\over m } g_{ij} {\dot \pi}^j . \eqno(1.9)$$
Using $(1.7)$ for ${\dot \pi}^j$ and taking onto account the symmetry of ${\dot x}_m {\dot x}_n$,
one obtains (with some rearrangement of indices) the geodesic equation
$$ {\ddot x}_\ell = - {\Gamma_\ell}^{mn}{\dot x}_m {\dot x}_n,
\eqno(1.10)$$
where the coefficients have the structure of the usual connection form
(here, $g^{ij}$ is the inverse of $g_{ij}$)
$$ \Gamma_\ell^{mn} = {1\over 2} g_{\ell k} \bigl\{ {\partial g^{km} \over
\partial x_n}+  {\partial g^{k n} \over
\partial x_m}- {\partial g^{n m} \over
\partial x_k} \bigr\}. \eqno(1.11)$$
This connection form is {\it compatible} with the metric $g_{ij}(x)$
by construction, {\it i.e.}, the covariant derivative constructed with
the $\Gamma_\ell^{mn}$ of $(1.10)$ of $g_{ij}$ vanishes, and we recognize that the dynamics
generated on the manifold $\{x\}$ is a geodesic flow.  It can carry,
moreover, a tensor structure which may be inferred from the invariance of the form $(1.2)$ under
local coordinate transformations (to be discussed below). In this framework, the definitions
$(1.3)$ and $(1.4)$ correspond to a special choice of coordinates.
\par However, this geodesic flow does not precisely follow the form of the
orbits generated by $(1.1)$ when the special choice of coordinates is used
for which $(1.3)$ and $(1.4)$ are valid. It has been shown that
in the conformal case which we are considering, a change in scale
along the orbits suffices to bring the motions into
correspondence [9] and achieve the stability criterion of
[1] within a single manifold. For many physical applications,
however, it is important to have both sets of variables
available (see, for
example, [10] and [11] in
application  to the theory of Milgrom[12], Bekenstein [13] and Bekenstein and Sanders[14]).
\par As we shall show,  the requirement that the momenta $p^i(t)$ and $\pi^i(t)$ are equal for all times is sufficient to determine the function $\phi(x)$ in an
open analytic set in terms of nonlinear functions of $F(q)$ and its
derivatives, and conversely, if $\phi(x)$ is known, to determine
$F(q)$ (and therefore $V(q)$), in an open analytic set, by the same
method. The existence of these analytic expansions demonstrates the
coexistence and correspondence of the coordinates on the two manifolds.
\par The relation
$$ {\dot x}_i = g_{ij}{\dot q}^j, \eqno(1.12)$$
as we shall show in Section 3, follows immediately from the requirement of dynamical equivalence of the two pictures. \footnote{*}{This relation was {\it assumed} in [1] in order to establish a correspondence between the Hamilton and geometric pictures}
The relation
$$dx_i = g_{ij}dq^j, \eqno(1.13) $$
which could be considered as implied by $(1.12)$,
 is clearly not integrable, so that one may not directly construct a global one to one correspondence between the two manifolds using this relation. One may, however, integrate it along a smooth geodesic curve, a procedure which will be adequate for our purposes.
Substituting $(1.12)$ into the geodesic formula $(1.11)$, one obtains the geodesic type relation
$$  {\ddot q}^\ell = -M^\ell_{mn}{\dot q}^m {\dot q}^n, \eqno(1.14)$$
where
$$ M^\ell_{mn}\equiv {1 \over 2}{\partial g_{n m} \over
\partial x_\ell}. \eqno(1.15)$$
\par As we have remarked above, the coordinates $\{x\}$ may carry local
diffeomorphisms;  ${\dot x}_j$ transforms (as we discuss in Section 2) under these
diffeomorphisms as a (contravariant) vector. In the special set of coordinates for which $(1.4)$ holds, the relation $(1.14)$ reduces to precisely the Hamilton equation
 $${\ddot q}^i= -{ 1 \over m} {\partial V(q)\over \partial q^i}, \eqno(1.16)$$
Therefore, the result $(1.14)$ constitutes a {\it geometric embedding} of the original Hamiltonian motion into a manifold, which we shall call the {\it Hamiltonian manifold} $\{q\}$, to which we assign, for simplicity in notation, the same symbol as the coordinates for the original Hamiltonian since it is a direct embedding (a special choice of coordinates for which $(1.4)$ is valid). As was shown in [1] and subsequent studies [4], the computation of the geodesic deviation of the orbits described by $(1.14)$ are remarkably predictive of instability in the original Hamiltonian motion, and have provided a new and effective criterion for the occurrence of such instabilities.  Before demonstrating the relation of the functions $\phi(x)$ and $F(q)$, we prove in the next section that the quantity $M^\ell_{ij}$ is a valid connection form, {\it i.e.}, under local diffeomorphisms, it provides the proper compensation terms to insure that $(1.14)$ is a covariant relation.
\bigskip
\noindent{\bf 2. Covariance of the embedded Hamiltonian motion}
\bigskip
\par In this section, we study the covariance of the dynamical equation representing the embedded  Hamiltonian motion.
\par It follows from the Hamiltion equations for the geometrical Hamiltonian
$(1.2)$ that
$$ {\dot x}_i = g_{ij}{\pi^j \over m}, \eqno(2.1)$$
or, using the assumption that $g_{ij}$ is invertible with inverse $g^{ij}$,
$$ \pi^j = m g^{ji}{\dot x}_i. \eqno(2.2)$$
Substituting this back into the Hamiltonian, one obtains
$$ H_G = {m \over 2} g^{ij}(x){\dot x}_i{\dot x}_j \eqno(2.3)$$
This form is srongly suggestive of the existence of a local diffeomorphism invariance for the manifold $\{x \}$ of the (contravariant) form\footnote{**}{This definition is inverse to that occurring in many standard texts, but corresponds, as required, to the transformation law of the differential $dx_i$, generally defined as contravariant.}
$$ dx'_i = {\partial x'_i \over \partial x_j} dx_j \eqno(2.4)$$
Requiring $H_G$ to be invariant under such diffeomorphisms, we see that the metric must transform as
$$ g'^{k\ell} = {\partial x_i \over \partial x'_k}{\partial x_j \over \partial x'_\ell} g^{ij}. \eqno(2.5)$$
\par We remark that from the result $(2.2)$ and the transformation property $(2.4)$, one obtains (with inverse Jacobian)
$$\pi'^j = {\partial x_\ell \over \partial x'_j}\pi^\ell, \eqno(2.6)$$
from which the inverse metric can be easily seen to transform as
$$ g'_{kj} = {\partial x'_k \over \partial x_m}{\partial x'_j \over \partial x_\ell} g_{\ell m}. \eqno(2.7)$$

 \par It follows from $(1.9)$, and the transformation law $(2.5)$  that
there is a diffeomorphism induced on the manifold $\{ q\}$ of the form
$$ dq'^j =  {\partial x_m \over \partial x'_j}dq^m, \eqno(2.8)$$
which appears to be covariant from the point of view of the $\{ x\}$ manifold, but is contravariant on $\{ q\}$.
\par We now study the covariance of Eq. $(1.14)$, writing it as
$$  {\ddot q}'^j = -M'^j_{k\ell}{\dot q}'^k {\dot q}'^\ell, \eqno(2.9)$$
where
$$ M'^j_{k\ell}\equiv {1 \over 2}{\partial g'_{k\ell} \over
\partial x'_j}. \eqno(2.10)$$
Formally dividing  $(2.8)$ by $dt$ on both sides and differentiating with respedct to $t$, one obtains
$$  {\ddot q}'^j = {\partial x'_p \over \partial x_k} {\partial^2 x_m \over \partial x'_j \partial x'_p} {\dot x}_k {\dot q}^m + {\partial x_m \over \partial x'_j} {\ddot q}^m . \eqno(2.11)$$
Clearly, ${\ddot q}^m$ does not transform as a tensor; such a transformation law would involve only the last term of $(2.11)$. We now show that the connection form $M'^j_{k\ell}$ develops a compensating term which returns the geodesic equation to its form before the transformation, up to a factor $ {\partial x_m \over \partial x'_j}$, {\it i.e.}, the diffeomorphism induced by the transformation.
Replacing ${\dot x}_k$ by $g_{k\ell} {\dot q}^\ell$ in the first term on the right, we have
$$ {\ddot q}'^j =g_{k\ell} {\partial x'_p \over \partial x_k} {\partial^2 x_m \over \partial x'_j \partial x'_p}{\dot q}^\ell {\dot q}^m + {\partial x_m \over \partial x'_j} {\ddot q}^m \eqno(2.12)$$
We now note that
$$ \eqalign{M'^j_{k\ell}&= {1\over 2} {\partial x_p \over \partial x'_j}
{\partial \over \partial x_p} ({\partial x'_k \over \partial x_m} g_{mn}
 {\partial x'_\ell \over \partial x_n})\cr
&= {1\over 2} {\partial x_p \over \partial x'_j}\bigl[ {\partial^2 x'_k \over \partial x_m \partial x_p} g_{mn}{\partial x'_\ell \over \partial x_n}\cr
&+{\partial x'_k \over \partial x_m}{\partial g_{mn} \over \partial x_p}      {\partial x'_\ell \over \partial x_n}\cr
&+ {\partial x'_k \over \partial x_m} g_{mn}{\partial^2 x'_\ell \over \partial x_n \partial x_p} \bigr].\cr} \eqno(2.13)$$
so  that
$$\eqalign{M'^j_{k\ell}{\dot q}'^k {\dot q}'^\ell &= {1\over 2} {\partial x_p \over \partial x'_j}\bigl[  2g_{m\ell}{\partial^2 x'_k \over \partial x_m \partial x_p} {\partial x_s \over \partial x'_k} \cr &+ {\partial g_{s\ell} \over \partial x_p}\bigr]{\dot q}^s {\dot q}^\ell.\cr}
\eqno(2.14)$$
It is, however, an identity, using
$$ {\partial  x'_p \over \partial x_k} {\partial \over \partial x'_j}
{\partial  x_m \over \partial x'_p} = - \bigl({\partial \over \partial x'_j}
{\partial  x'_p \over \partial x_k}\bigr){\partial  x_m \over \partial x'_p},
\eqno(2.15)$$
 that
$${\partial x_p \over \partial x'_j}g_{mq}{\partial^2 x'_\ell \over \partial x_m \partial x_p} {\partial x_s \over \partial x'_\ell}{\dot q}^s {\dot q}^q
 =-g_{k\ell} {\partial x'_p \over \partial x_k} {\partial^2 x_m \over \partial x'_j \partial x'_p} {\dot q}^\ell{\dot q}^m .
\eqno(2.16)$$
Then, we see that the first term of $(2.12)$ cancels with the addition of
  $ M'^j_{k\ell}{\dot q}^k {\dot q}^\ell$, so that
$$ {\ddot q}'^j +M'^j_{k\ell}{\dot q}'^k {\dot q}'^\ell
= {\partial x_m \over \partial x'_j}( {\ddot q}^m +M^m_{k\ell}{\dot q}^k {\dot q}^\ell),\eqno(2.17) $$
{\it i.e.}, the geodesic equation transforms as a tensor under this diffeomorphism. This expression correseponds to what we shall define as a {\it covariant derivative}, to be discussed further in Section 4.

 \par The connection $M^\ell_{mn}$ is, however, not
compatible with the metric $g_{ij}$, and therefore one cannot construct
a locally flat space ({\it i.e.}, an equivalence principle) in the Hamilton
framework, consistent with the fact that there is no equivalence principle in standard Hamilton mechanics.  Eq. $(1.10)$ is therefore, although covariant,
not a geodesic equation in the sense that it could be derived from a
minimum path length principle constructed with the metric $g_{ij}$. It
maintains, however,  the local diffeomorphism covariance derived from the
underlying Gutzwiller manifold, and therefore the resulting geodesic type equation corresponds to a proper
geometric embedding of the Hamiltonian motion.
\par  Although the dynamics represented by $(1.10)$ does not have an
equivalence
principle, parallel transport carried out in the Gutzwiller manifold
$\{x\}$, with transformation back to the Hamilton space $\{q\}$
results in precisely the truncated connection form $(1.11)$[1]. We shall see that this argument applies as well to the covariant derivative discussed in Section 4.
Applying
the method of geodesic deviation to the orbits described in terms of the geometric embedding
$(1.10)$  results in a stability criterion that has been found to be
remarkably effective in detecting chaotic behavior in Hamiltonian
systems [1],[4].
\par  We shall discuss the
nature of this embedding further in the next section.
\bigskip
\noindent{\bf 3. Relation between $\phi(x)$ and $F(q)$}
\bigskip
\par In this section we show systematically that all derivatives of the  function $\phi(x)$ are determined by $F(q)$ and its derivatives, and, conversely, that if $\phi(x)$ is known, all derivatives of $F(q)$ are determined by $\phi(x)$ by requiring that the momenta $p^i(t)$ and $\pi^i(t)$ are equal for all times. We shall call this requirement {\it dynamical equivalence}.  In a sufficiently analytic domain, the two functions can therefore be put into correspondence.
\par In the following, we shall use the explicit form of the metric for the geometric Hamiltonian in the special coordinate choice for which $(1.4)$ is valid, and for which\footnote{\S}{In this special choice of coordinates, the placement of indices does not always make the covariance evident.}
$$H_G = \phi(x) {\pi^2\over 2m}. \eqno(3.1)$$
\par Since
$$ V(q) = E( 1- {1 \over F}) \eqno(3.2)$$
the Hamilton equation for the time derivative of the momentum is
$${\dot p}^j = - {\partial V \over \partial q^j} =
 E {\partial \over \partial q^j} \bigl({ 1\over F} \bigr). \eqno(3.3)$$
\par For the geometric dynamics,
$$ \eqalign{{\dot \pi}^j &= - {\partial \phi \over \partial x_j}  {\pi^2\over 2m}=
 - {\partial \phi \over \partial x_j} {E\over \phi}\cr
&= -E {\partial \over \partial x_j} \ln \phi .\cr} \eqno(3.4)$$
\par It follows from $(3.3)$ and $(3.4)$ that (using the fact that $\phi = F$)
$$ \phi {\partial \phi \over \partial x_j} = {\partial F \over \partial q^j}. \eqno(3.5)$$
\par For the time derivatives of the coordinates, we  have, setting $p^j = \pi^j$,
$$ {\dot q}^j = {p^j \over m} = {\pi^j \over m} \eqno(3.6)$$
 and \footnote{*}{We remark that in this context, the conformal factor plays the role of a local mass scaling.}
$$ {\dot x}_j = {\phi \over m} \pi^j = {\phi \over m}p^j = \phi {\dot q}^j. \eqno(3.7)$$
We therefore see that the relation
$$ {\dot x}_i = g_{ij} {\dot q}^j, \eqno(3.8)$$
taken to be defined along a geodesic curve in $x^j$, is a necessary consequence of the dynamical equivalence of the two systems (see [9] for an alternative, but eventually equivalent, approach).
  \par Proceeding to the second derivatives, we have
$$ \eqalign{{\ddot p}^j &= E {\partial^2 \over \partial q^j \partial q^k }\bigl( {1 \over F}\bigr) {\dot q}^k\cr
{\ddot \pi} &= -E({\partial^2 \over \partial x_j \partial x_k }{\ln \phi}) {\dot x}^k .\cr} \eqno(3.9)$$
With $(2.1)$ and $(3.7)$ we may rewrite these equations as
$$ \eqalign{{\ddot p}^j &= E \bigl({\partial^2 \over \partial q^j \partial q^k } {1 \over F}\bigr) {\pi^k\over m} \cr
{\ddot \pi}^j &= -E\bigl({\partial^2 \over \partial x_j \partial x_k }{\ln \phi}\bigr) {\phi\over m} \pi^k\cr}. \eqno(3.10)$$
 Comparing coefficients of $\pi^k$, we find
$$ {\partial^2 \over \partial q^j \partial q^k }\bigl( {1 \over F}\bigr) =
- \bigl({\partial^2 \over \partial x_j \partial x_k} \ln \phi\bigr) \phi. \eqno(3.11)$$
 This result provides the equivalence to second order in the power series expansions of the functions $F(q)^{-1}$ and $\ln \phi$.
\par Explicitly, carrying out the derivatives, one obtains
$$ {2\over F^3} {\partial F \over \partial q^j}{\partial F \over \partial q^k} - {1 \over F^2}{\partial^2 F\over \partial q^j \partial q^k} = {1 \over \phi} {\partial \phi \over \partial x_j}{\partial \phi \over \partial x_k} - {\partial^2 \phi \over \partial x_j \partial x_k}. \eqno(3.12)$$
\par Substituting $(3.5)$ into the the first term on the right hand side, one obtains one half of the first term on the left; after cancellation, one obtains
$${\partial^2 \phi \over \partial x_j \partial x_k}= {1 \over F^2}\bigl({\partial^2 F\over \partial q^j \partial q^k} - {1\over F} {\partial F \over \partial q^j}{\partial F \over \partial q^k}\bigr). \eqno(3.13)$$
 \par This procedure can evidently be continued, and we therefore conclude that, in a domain of mutual analyticity, the Taylor series expansions of the two functions can be put into correspondence. We derive in the following a general formula.
\par It is convenient to write $(3.9)$ in the form
$$ \eqalign{{\ddot p}^j &= {E\over m} {\partial^2 \over \partial q^j \partial q^k }\bigl( {1 \over F}\bigr) p^k\cr
{\ddot \pi} &= -{E\over m}\bigl({\partial^2 \over \partial x_j \partial x_k }{\ln \phi}\bigr) \phi p^k. \cr} \eqno(3.14)$$
Differentiating with respect to $t$, one obtains
$$\eqalign{ ({d\over dt})^3p^j &= {E\over m} \bigl[{\partial^2 \over \partial q^j \partial q^k }\bigl( {1 \over F}\bigr)\bigr] p^k + {E \over m}{\partial^2 \over \partial q^j \partial q^k }\bigl( {1 \over F}\bigr) {\dot p}^k\cr
({d\over dt})^3\pi^j &= - {E \over m} {d \over dt} \bigl( {\partial^2 \over \partial  x_j \partial x_k} \ln \phi \bigr) \phi p^k - {E \over m}\bigl( {\partial^2 \over \partial  x_j \partial x_k} \ln \phi \bigr) \phi {\dot p}^k.\cr} \eqno(3.15)$$
Subtracting the two expressions, the last terms in each cancel by $(3.11)$, so that we have, carrying out the derivatives in $t$ to obtain factors of ${\dot q}^\ell = {p^\ell\over m}$ and ${\dot x}_\ell = \phi {p^\ell \over m}$,
$$ \eqalign{0 = ({d\over dt})^3p^j - ({d\over dt})^3\pi^j &= {E \over m^2} p^k p^\ell
\bigl[{\partial^3 \over \partial q^j \partial q^k \partial q^\ell} \bigl( {1\over F}\bigr)\cr &+\phi  {\partial\over \partial x_\ell} (\phi {\partial^2 \over \partial x_j \partial x_k} \ln \phi)\bigr].\cr} \eqno(3.16)$$
We therefore have the condition
$${\partial^3 \over \partial q^j \partial q^k \partial q^\ell} \bigl( {1\over F}\bigr) = -\phi  Sym {\partial\over \partial x_\ell} (\phi {\partial^2 \over \partial x_j \partial x_k} \ln \phi),\eqno(3.17)$$
where $Sym$ implies symmetrization.
\par Taking further derivatives with respect to $t$ of $(3.16)$ provides higher derivatives of the condition $(3.17)$ (with symmetrization), since derivatives of the momenta do not contribute due to the vanishing of the remaining factor (as sufficient conditions).  One then finds the condition for the fourth order that
$$ {\partial^4 \over \partial q^j\partial q^k\partial q^\ell\partial q^m} \bigl( {1\over F}\bigr)=
-\phi Sym {\partial \over \partial x_m }(\phi {\partial \over \partial x_\ell}(\phi {\partial^2 \over \partial x_j \partial x_k} \ln \phi)). \eqno(3.18)$$
It is clear that higher derivatives then obey the relation (as can easily be proved by induction)
$$ {\partial^n \over \partial q^{i_1}\partial q^{i_2}\dots \partial q^{i_n}} \bigl( {1\over F}\bigr)=
-\phi Sym {\partial \over \partial x_{i_n}} (\phi {\partial\over \partial x_{i_{n-1}}}(\dots \phi {\partial\over x_{i_3}}(\phi {\partial^2 \over \partial x_{i_2} \partial x_{i_1}} \ln \phi)\dots )). \eqno(3.19)$$
\par These relations between the coefficients of the Taylor expansion in the neighborhood of a given point in $x$ and $q$, are not, in general, integrable, but provide a relation between the functions $\phi(x)$ and $F(q)$ in analytic domains. We emphasize that the differentiability in $t$ that we have used was restricted to geodesic curves on $x$. We remark that the process may alternatively be carried out to provide formulas for the derivatives to all orders of $\phi(x)$ in terms of the function $F(q)$ and its derivatives.
\par It follows directly from this analysis that the Poisson bracket structure on the two manifolds is identical. To see this, we start with the time derivative of a function $f(x,\pi)$ on the geometrical manifold:
$${d \over dt} f(x,\pi) = {\partial f \over \partial x_i} {\dot x}_i + {\partial f \over \partial \pi^i} {\dot \pi}^i \eqno(3.20)$$
Since we are assuming Hamilton equations for the geometrical Hamiltonian $H_G$, this result is a Poisson bracket with the usual symplectic symmetry:
$$ {d \over dt} f(x,\pi)= \{ f, H_G\}. \eqno(3.21)$$
\par On the other hand, from $(2.1)$, the Poisson bracket $(3.20)$ becomes
$${d \over dt} f(x,\pi) = {\partial f \over \partial x_i}g_{ij}{\pi^j\over m}
 + {\partial f \over \partial \pi^i} {\dot \pi}^i \eqno(3.22)$$
 Since the Poisson bracket is formed by following the flow in phase space, the motion in $x_i$ follows a geodesic; therefore
${\partial f \over \partial x_i}$ is projected by ${\dot x}_i$ along a geodesic curve, and we can set
$${\partial f \over \partial x_i}g_{ij}= {\partial {\hat f} \over \partial q^j} \eqno(3.23)$$
in Eq. $(3.22)$,   where ${\hat f}$ is equal to $f$ considered a function of $q,p$.  Furthermore, $ {\pi^j\over m} = {p^j\over m}= {\dot q}^j$, so that $(3.22)$ becomes, with our requirement that the momenta $p$ and $\pi$ and all derivatives are equal,
$${d \over dt} f(x,\pi) = {\partial {\hat f} \over \partial q^i}{\dot q}^i
 + {\partial {\hat f} \over \partial p^i} {\dot p}^i .\eqno(3.24)$$
Therefore, we have
$${d \over dt} f(x,\pi) =
    {d \over dt} {\hat f}(q,p),     \eqno(3.25)$$
 establishing the equivalence of the Poisson brackets in both representations.
\bigskip
\noindent{\bf 4. Covariant derivatives on $\{q\}$}
 \bigskip
\par We have shown in the above that a geometric Hamiltonian of the form $(1.2)$ with conformal metric can be constructed which has properties closely related to the original Hamiltonian motion. The geodesics of the geometrical Hamiltonian $H_G$, in the special coordinate system for which the correspondence $(1.4)$ is constructed, do not follow the orbits of the original Hamiltonian, but the local mapping $(1.4)$ along the geodesic curve, required by the dynamical equivalence of the two pictures provides the modified geodesic type motion $(1.14)$ which, in this special choice of coordinates, precisely coincides with the equations of motion generated by the original Hamiltonian, as in $(1.16)$. We have, moreover, shown that the modified connection form $(1.15)$ is a good connection form, in the sense that the diffeomorphisms induced on the manifold denoted by $\{q\}$ by diffeomorphisms on $\{x\}$ leave the geodesic type equation $(1.14)$ invariant in form. In fact, Eq.$(1.14)$ can be written as the vanishing of a covariant derivative
$$ {D \over Dt} {\dot q}^k = \bigl( {d \over dt}{\delta^k}_j +{M^k}_{ij} {\dot q}^i \bigr) {\dot q}^j =0. \eqno(4.1)$$
\par The notion of covariant derivative in terms of parallel transport may be considered as parallel transport in the manifold $\{x\}$ and transformation back to the manifold $\{q\}$ in two steps, first from the tangent space of the geometric manifold to the manifold $\{x\}$, and then to $\{q\}$ [1]. The vanishing of the covariant derivative $(4.1)$ of ${\dot q}^k$ carries this geometrical interpretation, {\it i.e.}, it effectively carries the motion along the tangent space of $\{x\}$.  We would therefore expect that the second covariant derivative of ${\dot q}^k$ , and in fact, all covariant derivatives, would vanish as well.
\par  It is straightforward to compute the second covariant derivative:
$$ \eqalign{{D^2 \over Dt^2 }{\dot q}^k &= ({d \over dt})^2 {\dot q}^k + {3 \over 2} {\partial g_{m\ell} \over \partial x_k} {\ddot q}^m {\dot q}^\ell \cr
&+ {1 \over 2}  {\dot q}^p{\dot q}^m{\dot q}^\ell g_{qp} {\partial^2 g_{\ell m}\over \partial x_k \partial x_q} \cr
& + {1 \over 4}  {\partial g_{ij} \over \partial x_k} {\partial g_{m\ell} \over \partial x_j}{\dot q}^i{\dot q}^m {\dot q}^\ell, \cr} \eqno(4.2)$$
where we have used the replacement ${\dot x}_q = g_{pq} {\dot q}^p$ along the geodesic curve. In the special choice of coordinates, one then obtains
$${D^2 \over Dt^2 }{\dot q}^k= ({d \over dt})^2{\dot q}^k -{E\over m} \bigl\{{1 \over \phi} {\partial \phi \over \partial x_k }{ \partial \phi \over \partial x_\ell} - {\partial^2 \phi\over \partial x_k \partial x_\ell}\bigr\}{\dot q}^\ell. \eqno(4.3)$$
\par From $(3.9)$, which follows from equating the momenta, one may carry out the derivatives as in $(3.11)-(3.12)$ and divide by $m$ to show that the expression $(4.3)$ for the second covariant derivative vanishes. We infer that all covariant derivatives (for smooth motion) vanish, indicating that the
mechanism forming the geodesic motion on the geometrically embedded manifold $\{q\}$ is effectively associated with parallel transport on the geometric manifold $\{x\}$.
\par We emphasize that the vanishing of covariant derivatives is not the trivial result of differentiating a quantity that is identically zero; the covariant derivative is not an ordinary derivative, but contains information on the connection form so that the operation acts essentially on the tangent space. Since, in the geometrical embedding, the connection $M_{ij}^k$ is not compatible with the metric, this result demonstrates that the covariant derivative acts effectively, as we have pointed out before, on the tangent space of the underlying geometrical manifold.
\bigskip
\noindent{\bf 5. Variational correspondence}
\bigskip
\par In order to establish a variational orrespondence we now turn to the local relation between the functions $\phi(x)$ and $F(q)$. We wish to study the result of a variation in $q$ due to a variation in $x$ along a geodesic path. The relation imposed by dynamical equivalence $(1.8)$ is not sufficient to comprise a small but finite variation since for any small transport along a geodesic curve, the function $g_{ij}$ varies. We therefore define a variation $\delta q^i$  generated locally by the transport of a point $x$ along a geodesic as
$$ \delta q^i = \int_0^{\delta q}dq^i =  \int_0^{\delta \xi} g^{ij} (x_0 + \xi \eta) \eta_j d\xi, \eqno(5.1)$$
where we take $\xi$ to be an affine parameter along a small segment of the geodesic curve, which we approximate as a straight line with constant direction given by the unit vector ${\bf \eta}$. Since the upper limit on $\xi$ is small, we can expand in Taylor series to obtain
$$\eqalign{\delta q^i|_{x+\delta x} &= \int_0^{\delta \xi} \Bigl\{g^{ij}(x_0) \eta_j  +
{\partial g^{ij} \over \partial x_k} \eta_j  \eta_k \xi  \cr
&+ {1\over 2} {\partial^2 g^{ij}\over \partial x_k \partial x_\ell} \eta_j \eta_k \eta_\ell \xi^2 + \dots \Bigr\} d\xi; \cr} \eqno(5.2)$$
carrying out the integral over the affine paramter $\xi$, the resulting powers of $\delta \xi$  match the occurrence of the unit vectors, so we may write the result as
$$\eqalign{\delta q^i|_{x+\delta x} &=  g^{ij}(x_0) \delta x_j + {1\over 2}{\partial g^{ij} \over \partial x_k} \delta x_j\delta x_k \cr
&+ {1 \over 3!} {\partial^2 g^{ij}\over \partial x_k \partial x_\ell} \delta x_j\delta x_k\delta x_\ell + \dots \cr}\eqno(5.3)$$
\par We remark that since the derivative of $g^{ij}$ occurs symmetrically in $j,k$ in $(5.3)$, the second term is essentially complementary to the truncated connection $(1.11)$ with respect to a complete connection form of the type $(1.7)$. Thus the variation in $q^i$ is not along a geodesic curve in
 $\{ q\}$ corresponding to the geodesic curve in $\{x\}$ on which the variation is generated.
\par Formally, the structure continues to higher order (with the next term proportional to ${1 \over 4!}$)), but since we have approximated the geodesic curve by a straight line, we do not expect the result to be accurate to high orders; the second order term is sufficient for our present purposes. \footnote{\dag}{ Note that we could have followed the converse to define
$$\int_0^{\delta x} dx_j = \delta x_j|_{q +\delta q} = \int_0^{\delta q} g_{jk} (q_0 +\eta \xi) \eta^k d\xi,$$
and continue as in $(5.2)$. }
\par We then substitute this result into $F(q + \delta q)$; in the expansion of this function, one uses  $(5.3)$, {\it i.e.},
$$ \eqalign{F(q + \delta q)&= F( q +  g^{ij}(x_0) \delta x_j + {1\over 2}{\partial g^{ij} \over \partial x_k} \delta x_j\delta x_k \cr
&+ {1 \over 3!} {\partial^2 g^{ij}\over \partial x_k \partial x_\ell} \delta x_j\delta x_k\delta x_\ell + \dots) \cr
&= F(q) + {\partial F \over \partial q^i}(g^{ij}(x_0) \delta x_j + {1\over 2}{\partial g^{ij} \over \partial x_k} \delta x_j\delta x_k + \dots) \cr
&+{1 \over 2}{\partial^2 F \over \partial q^i \partial q^m}(g^{ij}(x_0) \delta x_j + {1\over 2}{\partial g^{ij} \over \partial x_k} \delta x_j\delta x_k + \dots) \cr
 &(g^{mj}(x_0) \delta x_j + {1\over 2}{\partial g^{mj} \over \partial x_k} \delta x_j\delta x_k + \dots) \cr
&+\dots \cr}\eqno(5.4)$$
 Comparison of the first order term with the expansion of $\phi(x +\delta x)$ yields  (recalling that $g^{ij} = {1 \over \phi}\delta _{ij}$  and contracting the $\delta$'s)  the relation $(3.5)$, which also follows from equating the first derivatives of the momenta $\pi$ and $p$.
 Keeping, moreover, second order terms from the first order part of the expansion, and adding them to the second order terms from the second order part of the expansion, one may compare with the second order part of the expansion of $\phi(x + \delta x)$. One finds (contracting the $\delta$'s) precisely the result
$(3.13)$.\footnote{\ddag}{Higher order terms in the expansion can easily be worked out, but since, as remarked above, we have assumed a locally linear form for the geodesic curve, their accuracy would be in doubt.}
\par  The second order result we have obtained is sufficient for our purposes.  What we have shown is that the functions $\phi(x)$ and $F(q)$ can be put locally into correspondence pointwise, establishing a relation between the manifolds $\{x\}$ and $\{q\}$ along geodesic curves; this relation is, furthermore, consistent with the condition of dynamical equivalence of the motions generated by the two Hamiltonians.  We have therefore established a basis for the correspondence between small variations in the two manifolds.
\bigskip
\noindent{\bf 6. Variational principles of Hamilton-Lagrange}
\bigskip
   \par In this section we recast the analytic mechanics of Hamilton
and Lagrange in the Hamilton space $\{q\}$ in terms of variations generated in an underlying geometrical manifold $\{x \}$. We start with the  Lagrangian obtained from the Legendre transform of $(1.2)$, {\it i.e.},
 $$ L_G = \pi^i {\dot x}_i -  g_{ij}(x) {\pi^i \pi^j \over 2m}. \eqno(6.1)$$
Since $\pi^i = g^{ij}{\dot x}/m$, one obtains
$$ L_G = { m \over 2}g^{ij} {\dot x}_i{\dot x}_j \eqno(6.2)$$
The variation of the action $S= \int dt L$ is then (leaving out the overall factor $m/2)$
$$ \delta S = \int dt \bigl\{ 2g^{ij} \delta {\dot x}_i{\dot x}_j + {\partial g^{ij} \over \partial x_k} \delta x_k {\dot x}_i{\dot x}_j \bigr\}. \eqno(6.3)$$
We now replace ${\dot x}_i$ by $g_{i\ell}{\dot q}^\ell$ and integrate by parts the time derivative on    $\delta {\dot x}_i$ to obtain an integral with coefficient $\delta x_i$:
$$\eqalign{\delta S &= \int dt \bigl\{ -2{\partial g^{ij} \over \partial x_m} g_{mn} g_{j\ell} {\dot q}^n {\dot q}^\ell  - 2 g^{ij} g_{mn}  {\partial g_{j\ell} \over \partial x_m}{\dot q}^n{\dot q}^\ell \cr
&- 2 g^{ij} g_{j\ell} {\ddot q}^\ell - {\partial g_{\ell m} \over \partial x_i} {\dot q}^\ell{\dot q}^m \bigr\}\delta x_i .\cr} \eqno(6.4)$$
Using the identity
$$ {\partial g^{ij} \over \partial x_m}= - g^{ip} {\partial g_{pq} \over \partial x_m}g^{qj}\eqno(6.5)$$
in the first term of $(6.4)$,
we see that the first two terms cancel; setting the coefficient of the arbitrary variation $\delta x_i$ to zero, one obtains the equation for the embedded Hamiltonian dynamics $(1.14)$.
\par The final form of the embedded Lagrangian can now be inferred, using our previous results, in the special coordinate system for which $(1.4)$ is valid, that is, writing (we delete an overall sign)
$$\delta S = \int dt \bigl\{ {\ddot q}^i + {1 \over 2} {\partial g_{\ell n} \over \partial x_i}{\dot q}^\ell {\dot q}^n\bigr\} \delta x_i \eqno(6.6)$$
and using $(1.3),(1.4)$ for $g_{\ell n}$, we find
$$\eqalign{ {\partial \phi \over \partial x_i } &= g^{ji}{\partial \phi \over \partial q^j }\cr
&= { 1 \over \phi}  {\partial \phi \over \partial q^j } \cr
&= {E-V \over E} {E \over (E-V)^2} {\partial V \over \partial q^i}\cr
&= {1 \over (E-V)} {\partial V \over \partial q^i}\cr}. \eqno(6.7)$$
The result $(6.4)$ then becomes, with ${\dot q}^2 = {2 \over m }(E-V)$,
$$ \delta S = \int dt \bigl\{ {\ddot q}^i + {1 \over m} {\partial V \over \partial q^i} \bigr\} \delta x_i, \eqno(6.8)$$
clearly equivalent to taking a Lagrangian of the form
 $$ L = { m \over 2} {\dot q}^2 -V(q). \eqno(6.9)$$
\par The variation of the associated action can be performed with respect to $\delta q^i = \delta x^i / \phi$, and is therefore arbitrary (we have assumed implicitly that $\delta x_i$ is along a geodesic curve; the arbitrariness of $\delta q_i$ then depends on the existence of a dense set of geodesics through the point  $x$). We have therefore recovered the standard Hamiltonian theory which was embedded in the manifold $\{q\}$.
\bigskip
\noindent{\bf 7. Summary and conclusions}
\bigskip
\par We have considered Hamiltonian mechanics on essentially three levels, the first, the standard potential model theory based on the Hamilton equations applied to a Hamiltonian $H$ of the form $(1.1)$, the second, an essentially equivalent geometric Hamiltonan $H_G$, and the third, a mapping to a geometric embedding of the original Hamiltonian motion.  While methods of computing Lyapunov exponents, and the application of the Jacobi metric (cited under[1]) are effective in many cases, there are notable exceptions, for example, in the analysis of Yahalom {\it et al} [5] of the work of ref. [15]. A method was developed in 2007 [1] involving the study of a geometric Hamiltonian [2],[3], which was put into dynamical equivalence with the standard Hamiltonian form by defining a conformal metric. On this level, one finds that the motion generated by the geometric Hamiltonian $H_G$ of $(1.2)$ can be transformed to a motion, defined by a connection form, on a geometric embedding of the original Hamiltonian motion.  In this representation, found in [1], the geodesic deviation results in a formula with remarkable predictive capability for the stability of the original Hamiltonian motion.
\par The mapping from the motion generated by $H_G$ to this geometric embedding, Eq. $(1.8)$, is a necessary condition for the dynamical equivalence (defined by setting $p^j (t) = \pi^j(t)$ for all $t$) of the two pictures, thus establishing the basis for the geometrical embedding.  This mapping cannot, clearly, be understood as a relation between closed one-forms, but only as a map in the tangent space along geodesic curves. There is therefore not a direct constructive global relation between the manifolds $\{q\}$ and $\{x\}$.  However, we have shown that the functions $\phi(x)$ and $F(q)= E /(E-V(q))$
are related in suitable analytic domains on $\{q\}$ and $\{x\}$ by well-defined relations between the coefficients of the Taylor series expansions of each of the functions around selected points $\{q\}$ and $\{x\}$. The Poisson bracket is, moreover, invariant under the mapping from the variables $x$ to $q$.
    \par  The Lagrangian formulation of the dynamics of these systems
makes explicit an intrinsic equivalence, thus accounting for the relation of the easily demonstrable instability (or stability) of the embedded motion to that induced by the original Hamiltonian.
\bigskip
\noindent{\bf 8. Acknowledgements}
\bigskip
\par We wish to thank Yossi Strauss, Michal Wagman, Gil Elgressy and Attay Kremer for helpful discussions.
\bigskip
\noindent{\bf 9. References}
\bigskip
\frenchspacing
\item {1.} L.P. Horwitz, Y. Ben Zion, M. Lewkowicz, M. Schiffer and J. Levitan, {\it Geometry of Hamiltonian Chaos}, Phys. Rev. Lett. {\bf 98} 23401 (2007). Geometrical methods of a different form were first introduced by C.G.J. Jacobi, {\it Vorlesungen \"uber Dynamik}, Verlag Reimer, Berlin 1884; J.S. Hadamard, {\it Les surfaces à courbures opposées et leurs lignes géodésiques}, J. Math. Pures Appl. {\bf 4}, 27 (1898), and further developed by L. Casetti and M. Pettini, {\it Analytic computation of the strong stochasticity threshold in Hamiltonian dynamics using Riemannian geometry}, Phys. Rev. E {\bf 48}, 4320 (1993); see also M. Pettini, {\it Geometery and Topology in Hamiltonian Dynamics and Statistical Mechanics}, Springer, New York (2006), and references therein.
\item{2.}M.C. Gutzwiller, {\it Chaos in Classical and Quantum Mechanics}, Springer-Verlag, new York (1990).
\item{3.}W.D. Curtis and F.R. Miller, {\it Differentiable Manifolds and Theoretical Physics}, Academic Press, New York (1985).
\item{4.}Y. Ben Zion and L.P. Horwitz, {\it Detecting order and chaos in three-dimensional Hamiltonian systems by geometrical methods}, Phys. Rev. E {\bf 76}, 046220 (2007); {\it Applications of geometrical criteria for transition to Hamiltonian chaos}, Phys. Rev. E {\bf 78} 036209 (2008); {\it Controlling Effect of Geometrically Defined Local Structural Changes on Chaotic Hamiltonian Systems}, Phys. Rev. E {\bf 81}, 046217 (2010).
\item{5.}A. Yahalom, J. Levitan and M. Lewkowicz, {\it Lyapunov vs. Geometrical Stability Analysis of  the Kepler and the Restricted Three-Body Problems}, Phys. Lett. A {\bf 375} 2111 (2011). See also, J. Levitan, A. Yahalom, L. Horwitz and M. Lewkowicz, {\it On the stability of Hamiltonian systems with weakly time dependent potentials}, Chaos: an Interdisciplinary Journal of Nonlinear Science {\bf 23}, 023122 (2013).
\item{6.}Lawrence Horwitz, Asher Yahalom, Meir Lewkowicz and Jacob Levitan, {\it Subtle is the Lord: On the difference between Hamiltonian (Lyapunov)stability analysis and geometrical stability analysis of gravitational orbits}, International Journal of Modern Physics D {\bf 20}, 2787 (2011).
\item{7.}A. Yahalom, M. Lewkowicz, J. Levitan, G. Elgressy, L.P. Horwitz and Y. Ben Zion, {\it Uncertainty relations for chaos}, International Journal of Geometric Methods in Modern Physics, {\bf 12}, 1550093 (2015).
\item{8.}Y. Strauss, L.P. Horwitz, J. Levitan and A. Yahalom, {\it Quantum field theory of classically unstable Hamiltonian dynamics}, Jour. Math. Phys. {\bf 56} 072701 (2015).
\item{9.}E. Calderon, L. Horwitz, R. Kupferman and S. Shnider, {\it On the geometrical formulation of Hamiltonian dynamics}, Chaos {\bf 23} 013120 (2013).
\item{10.}A. Gershon and L.P. Horwitz, {\it  Kaluza-Klein theory as a dynamics in a dual geometry}, Jour. Math. Phys.  {\bf 50}, 102704 (2009).
\item{11.}L.P. Horwitz, A. Gershon and M. Schiffer, {\it Hamiltonian map to conformal modification of space-time metric: Kaluza-Klein and TeVeS}, Found. of Phys. {\bf 41}, 141 (2010).
\item{12.}M. Milgrom, {\it A modification of the Newtonian dynamics as a possible alternative to the hidden mass hypothesis}, Astrophys. Jour.  {\bf 270}, 365, 
    {\it A modification of the Newtonian dynamics: Implications for Galaxies},
    371,
    {\it A modification of the Newtonian dynamics: Implications for Galaxy Systems}, 
    384 (1983).
\item{13.}J.D. Bekenstein, {\it Relativistic gravitation theory for the MOND paradigm}, Phys. Rev. D{\bf 70}, 083509 (2004).
\item{14} J.D. Bekenstein and R.H. Sanders, {\it A Primer to Relativistic MOND Theory}, EAS Pub. Series {\bf 20}, 225 (2006).
\item{15.}H. Safaai, M. Hasan and G. Saadat, {\it On the prediction of chaos in the restricted three-body problem}, in {\it Understanding Complex Systems}, p.369, Springer, Berlin (2006)

\end